\documentclass[superscriptaddress,
 reprint,
 amssymb, amsmath,
 aps, prl 
]{revtex4-2}
\usepackage[utf8]{inputenc}
\usepackage[T1]{fontenc}
\usepackage{amsmath}
\usepackage{amsfonts}
\usepackage{amssymb}
\usepackage{graphicx}
\usepackage{bm}
\usepackage{grffile} 
\usepackage{xr} 

\setcitestyle{super}

\DeclareMathOperator{\erf}{erf}
\DeclareMathOperator{\erfc}{erfc}

\graphicspath{{figures/}}

\begin{document}

\title{Universal bifurcations to explosive synchronization for networks of coupled oscillators with higher-order interactions}

\author{Lauren~D. Smith}
 \email{lauren.smith@auckland.ac.nz}
 \affiliation{\mbox{Department of Mathematics, The University of Auckland, Auckland 1142, New Zealand}}
\author{Penghao Liu}
 \affiliation{\mbox{Department of Mathematics, The University of Auckland, Auckland 1142, New Zealand}}

\date{\today}

\begin{abstract}
We determine critical parameter sets for transitions from gradual to explosive synchronization in coupled oscillator networks with higher-order coupling using self-consistency analysis. We obtain analytic bifurcation values for generic symmetric natural frequency distributions. We show that non-synchronized, drifting, oscillators are non-negligible, and play a crucial role in bifurcation. As such, the entire natural frequency distribution must be accounted for, rather than just the shape at the center. We verify our results for Lorentzian and Gaussian distributed natural frequencies. 
\end{abstract}

\maketitle





Many natural phenomena and engineering applications can be described as networks of coupled oscillators. For example, the firing of neurons in the brain \cite{BhowmikShanahan12, BickEtAl2020} and the dynamics of power grids \cite{MachowskiEtAl2011, NishikawaMotter2015, FilatrellaEtAl08, SchaferYalcin2019}. Classically, only pairwise interactions were considered, but it has recently been brought to light that many real-world networks have higher-order interactions \cite{PetriEtAl2014, LordEtAl2016, BattistonEtAl2021}, such that nodes interact as triplets or quadruplets, or larger groups. Moreover, it has been shown that these higher-order interactions yield fundamentally different dynamics than can be achieved with only pairwise coupling \cite{MajhiEtAl2022, BattistonEtAl2020}. In particular, for coupled oscillators, higher-order interactions have been shown to generate abrupt, explosive, transitions to synchronization that do not generically occur with pairwise coupling \cite{MillanEtAl2020, SkardalArenas2020, AdhikariEtAl2023, ZhangEtAl2023, GaoEtAl2023, BattistonEtAl2020, BattistonEtAl2021, MajhiEtAl2022}. These explosive transitions have been studied in detail for many classes of higher-order interactions, with most studies considering only Lorentzian distributed natural frequencies due to their amenability to the Ott-Antonsen ansatz \cite{OttAntonsen08,OttAntonsen09}. 

Here we determine critical parameter sets for general natural frequency distributions at which the onset of synchronization bifurcates from a gradual, smooth, transition, to an abrupt, explosive, transition. We do this using a self-consistency approach akin to that of Kuramoto's \cite{Kuramoto84, Strogatz00}. In Kuramoto's famous result, the onset of synchronization for symmetric frequency distributions is governed entirely by the shape of the frequency distribution at its center. We show that this is not the case for general higher-order dynamics because the population of ``drifters'' (oscillators that do not synchronize) cannot be neglected. As such, the entirety of the frequency distribution must be accounted for. We verify that our methodology recovers known formulae for Lorentzian distributed frequencies, and uncover analytic bifurcation values for Gaussian distributed frequencies.

\begin{figure*}[tbp]
\centering
\includegraphics[width=\textwidth]{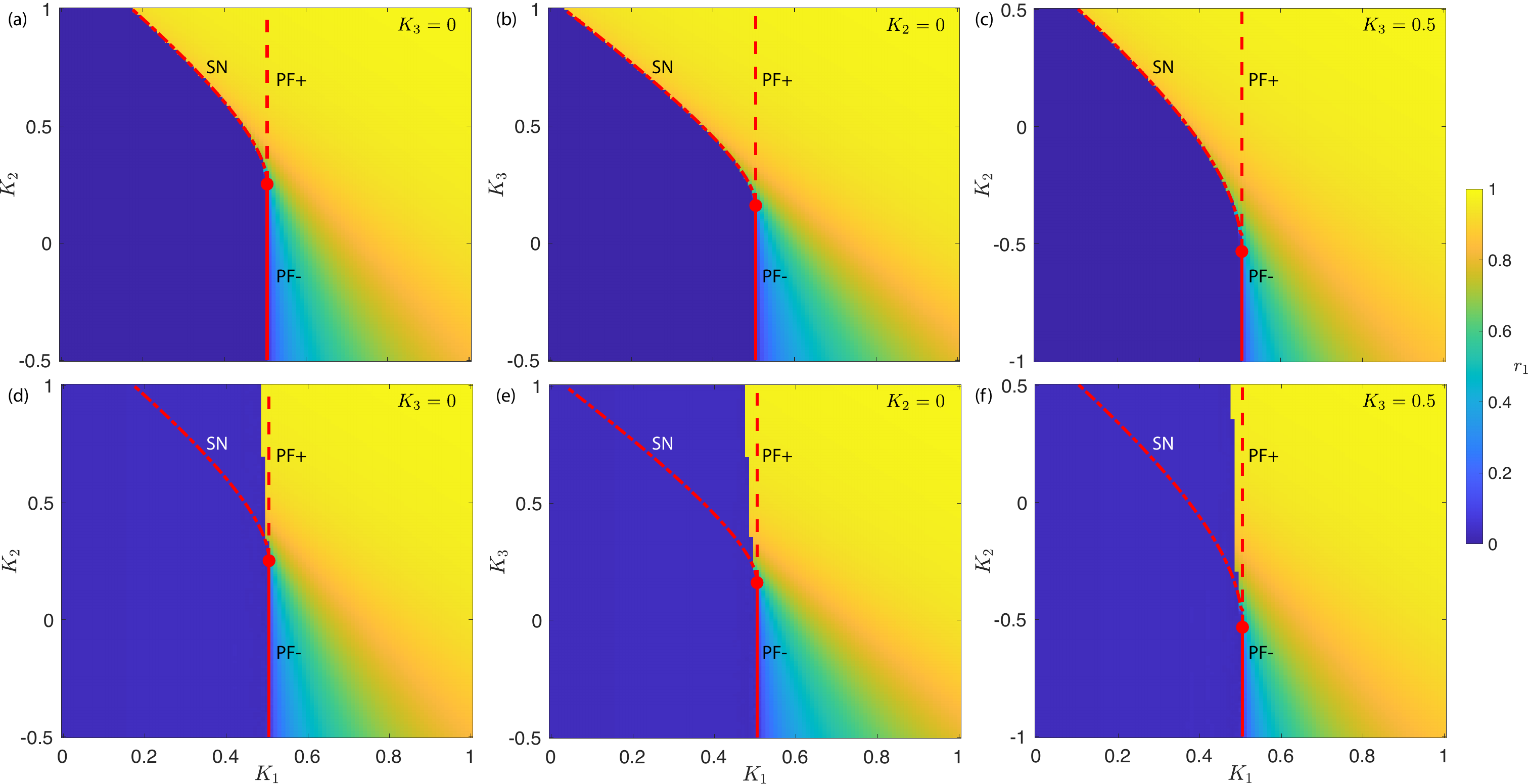}
\caption{
Transitions between explosive and gradual synchronization, quantified by the order parameter $r_1$, for the higher order system (\ref{eq:HO_KM}) with $N=10^4$ and a Gaussian natural frequency distribution $g(\omega)$ with mean zero and variance $\sigma^2 = 0.1$. Theoretical bifurcation values, from (\ref{eq:kc}) and (\ref{eq:crit_K2_K3}) are shown in red -- solid: supercritical pitchfork bifurcation (PF-); dashed: subcritical pitchfork bifurcation (PF+); dot-dashed: saddle-node bifurcation (SN). The bifurcation curves meet at a co-dimension two bifurcation point given by (\ref{eq:kc}) and (\ref{eq:crit_K2_K3}) (red circle). Top row: Highly synchronized initial condition. Bottom row: Uniformly random initial condition. (a), (d)~$K_3 = 0$, $K_1$, $K_2$ varying. (b), (e)~$K_2=0$, $K_1$, $K_3$ varying. (c), (f)~$K_3 = 0.5$, $K_1$, $K_2$ varying.
}
\label{fig:r1_combined}
\end{figure*}

We focus on the higher-order dynamics given by
\begin{align}
\dot{\theta}_i = \omega_i &+ \frac{K_1}{N} \sum_{j=1}^N \sin\left(\theta_j - \theta_i\right)\nonumber \\
& + \frac{K_2}{N^2} \sum_{j,l=1}^N \sin\left(2\theta_j - \theta_l - \theta_i\right) \label{eq:HO_KM} \\
&+ \frac{K_3}{N^3} \sum_{j,l,m=1}^N \sin\left(\theta_j + \theta_l - \theta_m - \theta_i\right), \nonumber
\end{align}
where $N$ is the number of oscillators, $\omega_i$ are natural frequencies that are drawn from a probability distribution $g(\omega)$, and $K_1$, $K_2$, $K_3$ are the coupling strengths of the dyadic, triadic and tetradic couplings, respectively. This model derives from the mean-field complex Ginzburg-Landau equation \cite{LeonPazo2019}, or from weakly coupled Hopf bifurcations \cite{AshwinRodrigues2016}. The model (\ref{eq:HO_KM}) has been studied in detail using the Ott-Antonsen approach in the limit $N\to \infty$ assuming Lorentzian distributed natural frequencies \cite{SkardalArenas2020}. It was shown that there is a bifurcation in the onset of synchronization from a gradual second-order transition to an explosive first-order transition. Here we use self-consistency to analyze these bifurcations for general symmetric natural frequency distributions $g(\omega)$. For instance, using a Gaussian distribution with mean zero and variance $\sigma^2 = 0.1$, Fig.~\ref{fig:r1_combined} shows bifurcations from gradual to explosive synchronization (quantified by the order parameter $r_1$ defined below) as the parameters $K_1,K_2,K_3$ are varied. We derive analytic expressions for the locations of these bifurcations.

Self-consistency has been used for subcases of the dynamics (\ref{eq:HO_KM}), for instance, using $K_1 = K_3 = 0$ \cite{WangEtAl2021}  (where the dynamics are fundamentally different: the unstable branch at onset of explosive synchronization never reconnects via a pitchfork bifurcation), using $K_2 = 0$ \cite{XuEtAl2023} (in which case drifters can be ignored), and using $K_3=0$ to study oscillators with inertia \cite{Sabhahit2023}, but so far the full system (\ref{eq:HO_KM}) has not been studied via self-consistency and universal bifurcation sets have not been found.


We follow a similar self-consistency approach to that of Kuramoto \cite{Kuramoto84}, which has been generalized for several classes of higher-order interactions \cite{WangEtAl2021, WangEtAl2021b, XuEtAl2020, XuEtAl2021, XuEtAl2023, Sabhahit2023} (though not for (\ref{eq:HO_KM})). In the limit $N\to \infty$, the oscillator population is defined by a density $\rho(\theta,\omega,t)$, where $\rho(\theta,\omega,t)d\theta d\omega$ is the fraction of oscillators with phases in the range $(\theta,\theta+d\theta)$ and frequencies in the range $(\omega,\omega+d\omega)$ at time $t$. We will determine stationary distributions $\rho(\theta,\omega)$. To do so we write the dynamics (\ref{eq:HO_KM}) in mean-field form 
\begin{align}
\dot{\theta}_i = \omega_i &+ K_1 r_1 \sin\left(\Psi_1 - \theta_i\right) + K_2 r_1 r_2 \sin\left(\Psi_2 - \Psi_1 - \theta_i\right) \nonumber \\ 
&+ K_3 r_1^3 \sin\left(\Psi_1 -\theta_i \right), \label{eq:HO_KM_MF}
\end{align}
where 
\begin{equation} \label{eq:order_params}
r_1 e^{i \Psi_1} = \frac{1}{N} \sum_{j=1}^N e^{i\theta_j}, \quad r_2 e^{i \Psi_2} =  \frac{1}{N} \sum_{j=1}^N e^{2 i\theta_j} ,
\end{equation}
denote the classical complex order parameter and a Daido order parameter \cite{Daido1996}, respectively.
Assuming $g(\omega)$ is symmetric, in the limit $N\to \infty$ we find $\Psi_2 = \Psi_1$, and we can assume, without loss of generality, that $\Psi_1 = 0$. Thus the mean-field form (\ref{eq:HO_KM_MF}) simplifies to
\begin{equation}
\dot{\theta}_i = \omega_i -\gamma \sin\theta_i,  \label{eq:HO_KM_MF_2}
\end{equation}
where $\gamma = r_1\left(K_1 + K_2 r_2 + K_3 r_1^2\right)$.
Equilibria of (\ref{eq:HO_KM_MF_2}) satisfy
\begin{equation} \label{eq:locked_eq} 
\omega_i = \gamma \sin \theta_i,
\end{equation} 
for oscillators satisfying $|\omega_i|<\gamma$. Such oscillators form the synchronized cluster, and are termed ``locked''. Thus, the stationary density satisfies
\begin{equation} \label{eq:locked_density}
\rho(\theta,\omega) = \delta(\omega - \gamma \sin \theta), \quad |\omega|<\gamma.
\end{equation}
 In partially synchronized states, there also exist ``drifters'' which do not synchronize and are those with frequencies $|\omega_i|>\gamma$. As in the Kuramoto model, stationarity of the drifter density requires that $\rho(\theta,\omega)$ is inversely proportional to the speed at $\theta$, i.e.,
\begin{equation} \label{eq:drifter_density}
\rho(\theta,\omega) = \frac{C(\omega,\gamma)}{|\omega - \gamma \sin \theta|}, \quad |\omega|>\gamma
\end{equation}
where $C(\omega,\gamma) =(2\pi)^{-1} \sqrt{\omega^2 - \gamma^2}$ is a normalization constant chosen such that $\int_{-\pi}^\pi \rho d\theta = 1$. The stationary density (\ref{eq:locked_density})-(\ref{eq:drifter_density}) shares a strong similarity to that of the original Kuramoto model \cite{Kuramoto84, Strogatz00}, except with $Kr_1$ replaced by $\gamma$. The symmetries $\rho(-\theta,-\omega) = \rho(\theta,\omega)$ and $\rho(\theta+\pi,-\omega) = \rho(\theta,\omega)$ follow readily, and are useful in our self-consistency analysis.

Self-consistency requires that the order parameters defined implicitly via (\ref{eq:locked_density})-(\ref{eq:drifter_density}) are consistent with the definitions in (\ref{eq:order_params}). Noting that $\Psi_1=\Psi_2 = 0$, the equations are self-consistent provided
\begin{equation} \label{eq:SC_all}
r_k =  r_k^l + r_k^d,\quad k=1,2,
\end{equation}
where $r_k^l$ and $r_k^d$ are the locked and drifter components of $r_k$, respectively.  As in the case of the Kuramoto model, the symmetries of $g(\omega)$ and $\rho(\theta,\omega)$ ensure that the imaginary parts of both the locked and drifter components are zero for all $k$. Focusing first on the locked component, using (\ref{eq:locked_density}) we obtain
\begin{align} 
r_k^l &= \int_{-\gamma}^\gamma \cos\left(k \theta(\omega)\right) g(\omega) d\omega \nonumber \\
&= \gamma \int_{-\pi/2}^{\pi/2} \cos(k\theta)\cos(\theta)\, g(\gamma \sin\theta) d\theta, \label{eq:SC_locked}
\end{align}
where $\theta(\omega)$ is defined implicitly via (\ref{eq:locked_eq}). For the drifter component, using (\ref{eq:drifter_density}) we obtain
\begin{equation}  \label{eq:SC_rk_d}
r_k^d = \int_{|\omega|>\gamma} \int_{-\pi}^\pi \cos(k\theta) g(\omega) \rho(\theta,\omega) d\theta d\omega.
\end{equation}
Using the symmetries of $g$ and $\rho$ it is easily shown that $r_1^d=0$ \footnote{See Appendix A for full detail.}, and, hence, $r_1 = r_1^l$. However, $r_2^d \neq 0$, in fact, after a change of variables $\eta = \omega/\gamma$, we obtain
\begin{equation} \label{eq:SC_r2_d}
r_2^d = 2\gamma \int_1^\infty I_\theta(\eta) g(\gamma \eta) d\eta,
\end{equation}
where, assuming $\eta>1$, 
\begin{equation} \nonumber
I_\theta(\eta) = \int_{-\pi}^\pi \cos(2\theta)\rho(\theta,\eta) d\theta = 1 + 2\eta\left(\sqrt{\eta^2 -1} - \eta\right).
\end{equation}
The combination of (\ref{eq:SC_all}), (\ref{eq:SC_locked}) and (\ref{eq:SC_r2_d}) form the self-consistency equations for the model (\ref{eq:HO_KM}). For given parameters $K_1,K_2,K_3$, the solutions to the self-consistency equations give the order parameters $r_1$ and $r_2$, and, hence, the stationary density (\ref{eq:locked_density})-(\ref{eq:drifter_density}).

Having derived the self-consistency equations, we use them to determine bifurcations in the type of synchronization at onset. We use the general principle described in Kuehn \& Bick \cite{KuehnBick2021} to determine when the onset of synchronization bifurcates from gradual to explosive. Considering the values of $K_2$ and $K_3$ to be fixed, with $K_1$ varying, the incoherent state $r_1=r_2=0$ is stable for small values of $K_1$, i.e., those less than some $K_c$, which may depend on $K_2,K_3$. At the critical value $K_1 = K_c$, a branch of non-zero solutions for $r_1,r_2$ emerges via a pitchfork bifurcation. If the pitchfork bifurcation is supercritical, i.e., the new solution branch exists for $K_1>K_c$, then the new branch is stable, indicating the onset of stable synchronization via a (gradual) second-order phase transition. Conversely, if the pitchfork bifurcation is subcritical, i.e., the new solution branch exists for $K_1<K_c$, then the new branch is unstable, which indicates an (explosive) first-order phase transition to synchronization must occur at $K_1 = K_c$, and the synchronized state will vanish via a saddle-node bifurcation at some value $K_1^{\rm{SN}}(K_2,K_3) < K_c$. In this case there is bistability in the parameter region $K_1^{\rm{SN}}(K_2,K_3)<K < K_c$, with both the incoherent and synchronized states being stable (cf. the top row compared to the bottom row in Fig.~\ref{fig:r1_combined}). Therefore, in order to determine when the onset of synchronization becomes explosive, it is sufficient to determine $K_c$, and whether the new non-zero branch of solutions for $r_1,r_2$ exists for $K_1>K_c$ or $K_1<K_c$. 

To determine $K_c$, we find the critical parameter at which the new non-zero branch emerges from the incoherent branch $r_1 = r_2 =0$. Therefore, we consider the limit as $r_1,r_2\to 0$ of solutions to (\ref{eq:SC_all}). For $r_1 = r_1^l$ (\ref{eq:SC_locked}), assuming $r_1 \neq 0$ we can divide both sides of (\ref{eq:SC_locked}) by $r_1$ to obtain
\begin{equation} \nonumber
1 = \left(K_1 + K_2 r_2\right) \int_{-\pi/2}^{\pi/2} \cos^2(\theta)\, g(\gamma \sin\theta) d\theta.
\end{equation}
In the limit $r_1,r_2\to 0$ we obtain $1 = K_1 \pi g(0)/2$, and, hence, the non-zero branch reaches zero at 
\begin{equation} \label{eq:kc}
K_1 = K_c = \frac{2}{\pi g(0)},
\end{equation}
which is identical to the critical coupling strength for the onset of synchronization in the classical Kuramoto model \cite{Kuramoto84}. While we previously noted that $K_c$ may depend on $K_2$ and $K_3$, we find that it is independent of $K_2$ and $K_3$. Instead, we will show that $K_2$ and $K_3$ control whether the non-zero branching solution is stable or unstable, and, hence, whether synchronization occurs gradually or abruptly.

We now determine the criticality of the pitchfork bifurcation at $K_1=K_c$, with $K_1$ varying and $K_2,K_3$ kept fixed. This will yield critical bifurcation parameters that separate explosive and gradual synchronization transitions.
To do this, we expand the self-consistency equations (\ref{eq:SC_all}), (\ref{eq:SC_locked}) and (\ref{eq:SC_r2_d}) in Taylor series about the critical point $(r_1,r_2,K_1) = (0,0,K_c)$, with $K_2$ and $K_3$ treated as constants. The right-hand-sides of the self-consistency equations (\ref{eq:SC_all}) are functions of $\gamma$, and so we first expand about $\gamma=0$ then substitute $\gamma =K_c  r_1  + r_1 (K_1 -K_c) + K_2 r_1 r_2 + K_3 r_1^3$. For the locked components (\ref{eq:SC_locked}) we obtain
\begin{align}
r_1 &= r_1^l = \frac{1}{K_c} \gamma + \frac{\pi g''(0)}{16} \gamma^3 + \mathcal{O}(\gamma^5), \label{eq:r1_TS} \\
r_2^l &= \frac{2 g(0)}{3} \gamma - \frac{g''(0)}{15} \gamma^3+ \mathcal{O}(\gamma^5). \label{eq:r2_l_TS}
\end{align}
For the drifter component of $r_2$ (\ref{eq:SC_r2_d}) we obtain
\begin{equation} \label{eq:r2_d_TS}
r_2^d = -\frac{2g(0)}{3} \gamma + c_2 \gamma^2 + \mathcal{O}(\gamma^3),
\end{equation}
where
\begin{equation} \label{eq:c_2}
c_2 = \frac{1}{2} \lim_{\gamma \to 0} \frac{d^2 r_2^d}{d\gamma^2} =2 \lim_{\gamma\to 0} \int_1^\infty \eta I_\theta(\eta) g'(\gamma\eta) d\eta.
\end{equation}
Here we have used the fact that the integral $\int_1^\infty I_\theta(\eta) d\eta = -1/3$ converges, which allows us to apply the dominated convergence theorem to evaluate $\lim_{\gamma\to 0} r_2^d=0$ and $\lim_{\gamma\to 0} \frac{dr_2^d}{d\gamma}=-\frac{2g(0)}{3}$. However, the integral $\int_1^\infty\eta I_\theta(\eta) d\eta$ diverges, meaning the limit and integral in (\ref{eq:c_2}) cannot be interchanged \footnote{See Appendix A for full details.}. Physically, this reflects that the contribution of the drifters is not localized to $\omega =0$, and the entirety of the natural frequency distribution $g(\omega)$ must be accounted for. We demonstrate this explicitly for two distributions with a common central region in Appendix B. Combining (\ref{eq:r2_l_TS}) and (\ref{eq:r2_d_TS}) we obtain the expansion
\begin{equation}
r_2 = c_2 \gamma^2 - \frac{g''(0)}{15} \gamma^3+ \mathcal{O}(\gamma^5). \label{eq:r2_TS}
\end{equation}
We now substitute $\gamma =K_c  r_1  + r_1 (K_1 -K_c) + K_2 r_1 r_2 + K_3 r_1^3$ into (\ref{eq:r1_TS}) and (\ref{eq:r2_TS}) and collect powers up to second order in $r_1$, $r_2$ and $(K_1-K_c)$, yielding
\begin{align}
0 &= \frac{1}{K_c} (K_1-K_c) + \frac{K_2}{K_c} r_2 +\left( \frac{K_3}{K_c} + \frac{\pi K_c^3 g''(0)}{16}\right) r_1^2, \nonumber  \\
r_2 &= c_2 K_c^2 r_1^2 , \nonumber 
\end{align}
where we have first divided (\ref{eq:r1_TS}) through by $r_1$. Solving these equations simultaneously gives
\begin{equation} \nonumber 
r_1^2 = -\frac{ 16(K_1 - K_c) }{ 16(c_2 K_c^2 K_2 + K_3) + \pi K_c^4 g''(0)}.
\end{equation}
From our previous discussion, the onset of synchronization will switch from gradual to explosive when this solution switches from existing to the right of $K_1 = K_c$ to existing to the left of $K_1 = K_c$, i.e., at the critical parameter values
\begin{equation} \label{eq:crit_K2_K3}
c_2 K_c^2 K_2 + K_3 = K_{2,3}^* = - \frac{\pi K_c^4 g''(0) }{16} .
\end{equation}
Assuming that $g''(0)<0$, which is true for unimodal distributions, the onset of synchronization is gradual for $c_2 K_c^2 K_2 + K_3 < K_{2,3}^*$, and explosive for $c_2 K_c^2 K_2 + K_3 > K_{2,3}^*$.

From (\ref{eq:crit_K2_K3}), if $K_2=0$, then the bifurcation is at $K_3 = K_{2,3}^*$, which does not depend on $c_2$, and only depends on $g(0)$ and $g''(0)$. This is similar to Kuramoto's result \cite{Kuramoto84}, in which bifurcation only depends on the shape of $g$ at $0$. The cause of this is that the drifters can be neglected if $K_2 =0$  \cite{XuEtAl2023}, whereas the drifters have a non-negligible effect, quantified by $c_2$, if $K_2 \neq 0$.



We now demonstrate our results explicitly, first for Lorentzian distributed natural frequencies, and then for Gaussian distributed natural frequencies. For a Lorentzian distribution centered at zero and with spread $\Delta = 1$ we have $g(\omega) = (\pi(1+\omega^2))^{-1}$. Evaluating the self-consistency integrals (\ref{eq:SC_locked}) and (\ref{eq:SC_r2_d}) directly yields
\begin{align}
r_1 & = \frac{\sqrt{1+\gamma^2} - 1}{\gamma}, \nonumber \\
r_2 &= \frac{2+\gamma^2 - 2\sqrt{1+\gamma^2}}{\gamma^2} = r_1^2. \nonumber 
\end{align} 
The first equation can be solved for $\gamma$ which yields $\gamma = 2r_1/(1-r_1^2)$. We also have our definition $\gamma = r1(K_1 + K_2 r_2 + K_3 r_1^2)$. Solving this system of equations for $r_1$ yields
\begin{equation} \label{eq:Lorentz_r1}
0 = r_1\left( -2 + K_1(1-r_1^2) + K_{2+3} r_1^2(1-r_1^2)\right),
\end{equation}
where $K_{2+3} = K_2 + K_3$, as defined in Skardal \& Arenas \cite{SkardalArenas2020}. Equation (\ref{eq:Lorentz_r1}) is identical to the equivalent equation obtained via the Ott-Antonsen approach \cite{SkardalArenas2020}, which has been shown to agree with the full dynamics (\ref{eq:HO_KM}) for large $N$. For the Lorentzian distribution we obtain $K_c = 2$ meaning that the new branch of non-zero solutions $r_1,r_2$ emanate from $K_1 = K_c = 2$. We also directly compute $c_2 = 1/4$ and $K_{2,3}^* = 2$, and, hence, our criticality condition (\ref{eq:crit_K2_K3}) becomes $K_{2+3} = K_{2,3}^* = 2$.
Our results agree with the analysis using the Ott-Antonsen approach \cite{SkardalArenas2020}, which finds a co-dimension two bifurcation at $(K_1,K_{2+3}) = (2,2)$, such that at bifurcation the pitchfork bifurcation changes criticality, and a saddle-node bifurcation emerges.

We now consider Gaussian distributed natural frequencies, with mean zero and variance $\sigma^2$ (in our numerical examples we use $\sigma^2=0.1$). We can again directly integrate the self-consistency integrals (\ref{eq:SC_locked}) and (\ref{eq:SC_r2_d}) to obtain
 \begin{align}
r_1 & = \frac{\sqrt{\pi}}{2} A e^{-A^2/2} \left(I_0(A^2/2) + I_1(A^2/2)\right)  , \label{eq:r1_Gaussian} \\
r_2 &= 1 + \frac{e^{-A^2} - 1}{ A^2},\label{eq:r2_Gaussian}
\end{align} 
where $A = \gamma/(\sqrt{2}\sigma)$ and $I_n$ denotes the $n$-th modified Bessel function of the first kind. In principle, (\ref{eq:r2_Gaussian}) can be solved for $A$, and then the result substituted into (\ref{eq:r1_Gaussian}), which can then be solved for $r_1$, however, that is not possible in practice, except numerically. We note that the relation $r_2 = r_1^2$ that was true for the Lorentzian distribution, and which allowed the simplification of $K_2$ and $K_3$ to the combined variable $K_{2+3}$, is not true for Gaussian distributions. This is reflected in Fig.~\ref{fig:r1_combined} by the differences between Fig.~\ref{fig:r1_combined}(a,d) and Fig.~\ref{fig:r1_combined}(b,e). For our criticality results, we find $K_c = 2\sigma \sqrt{2/\pi}$, $c_2 = 1/(4\sigma^2)$ and $K_{2,3}^* = 2\sigma \sqrt{2/\pi^3}$. The red curves in Fig.~\ref{fig:r1_combined} show that these theoretical values agree well with numerical simulations of the full model (\ref{eq:HO_KM}). Our numerical results show the onset of synchronization from a random initial condition (bottom row) occurs at $K_1 = K_c = 0.5046$ via a pitchfork bifurcation, with a switch from gradual to explosive synchronization as $K_2$ or $K_3$ are increased, corresponding to the change in criticality of the pitchfork bifurcation (shown as solid for supercritical (PF-) and dashed for subcritical (PF+)). From (\ref{eq:kc}) and (\ref{eq:crit_K2_K3}), for $K_3=0$ (Fig.~\ref{fig:r1_combined}(a,d)) this co-dimension two bifurcation occurs at $(K_1,K_2) = (K_c,K_{2,3}^*/(c_2 K_c^2)) = (0.5046, 0.2523)$. For $K_2=0$ (Fig.~\ref{fig:r1_combined}(b,e)) the co-dimension two bifurcation occurs at $(K_1,K_3) = (K_c,K_{2,3}^*) = (0.5046, 0.1606)$, and for $K_3 = 0.5$ (Fig.~\ref{fig:r1_combined}(c,f)) the co-dimension two bifurcation occurs at $(K_1,K_2) = (K_c,(K_{2,3}^*-0.5)/(c_2 K_c^2)) = (0.5046, -0.5331)$

\begin{figure}[tbp]
\centering
\includegraphics[width=\columnwidth]{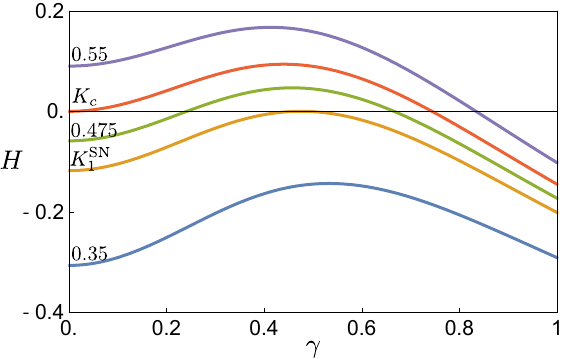}
\caption{
$H(\gamma)$ given by (\ref{eq:H_of_gamma}) for Gaussian distributed natural frequencies with mean zero and variance $\sigma^2 = 0.1$ using $K_2 = 0.5$, $K_3=0$ and a range of $K_1$ values (labeled on the figure).
}
\label{fig:H_of_gamma}
\end{figure}

When the onset to synchronization is explosive, i.e., when $c_2 K_c^2 K_2 + K_3 > K_{2,3}^*$, the branch of non-zero solutions for $r_1,r_2$ that emanate from $K_1 = K_c$ are unstable, and there is a saddle-node bifurcation that occurs at some $K_1^{\rm{SN}}(K_2,K_3)$ such that the unstable branch meets a stable branch of synchronized solutions. For Lorentzian distributed natural frequencies the saddle-node bifurcation can be found easily by directly solving (\ref{eq:Lorentz_r1}) \cite{SkardalArenas2020}. However, for Gaussian distributed natural frequencies the same approach would require analytically solving (\ref{eq:r1_Gaussian})-(\ref{eq:r2_Gaussian}), which is not possible. As a method to (numerically) find the saddle-node bifurcation for general frequency distributions, we recognize that our self-consistency equations (\ref{eq:SC_all})-(\ref{eq:SC_r2_d}) are of the form $r_1 = r_1(\gamma) = \gamma F_1(\gamma)$ and $r_2 = r_2 (\gamma)$. From the definition of $\gamma$, this sets up a single self-consistency equation for $\gamma$:
\begin{equation} \nonumber 
\gamma = r_1(\gamma)\left( K_1 + K_2 r_2(\gamma) + K_3 r_1(\gamma)^2\right)
\end{equation}
which has solutions $\gamma = 0$ and $\gamma$ satisfying
\begin{equation} \label{eq:H_of_gamma}
0 =  F_1(\gamma) \left(K_1 + K_2 r_2(\gamma) + K_3 \gamma^2 F_1(\gamma)^2\right)-1  = H(\gamma) .
\end{equation}
Each positive root of $H(\gamma)$ corresponds to a solution branch of the self-consistency equations (\ref{eq:SC_all})-(\ref{eq:SC_r2_d}). Thus, finding the saddle-node bifurcation is equivalent to finding double-roots of $H$, i.e., solutions to $H(\gamma) = 0$ and $H'(\gamma)=0$. For Gaussian natural frequencies $H(\gamma)$ is obtained from (\ref{eq:r1_Gaussian})-(\ref{eq:r2_Gaussian}), and shown in Fig.~\ref{fig:H_of_gamma} for a range of $K_1$ values with $K_2=0.5$ and $K_3=0$ kept fixed.
At $K_1=0.35$ there are no roots, and, hence, no synchronized solution. Increasing $K_1$, at $K_1^{\rm{SN}} = 0.445$ $H(\gamma)$ has a double-root, indicating a saddle-node bifurcation and birth of a synchronized state via an explosive transition. For $ K_1^{\rm{SN}}<K_1<K_c = 0.505$, $H(\gamma)$ has two roots: one stable and one unstable synchronized solution. For $K_1 >K_c$, there is a single root of $H(\gamma)$; the unstable solution has been destroyed in the pitchfork bifurcation at $K_1 = K_c$.
We have numerically found the saddle-node bifurcation, $K_1^{\rm{SN}}(K_2,K_3)$, for Gaussian natural frequencies for a range of $K_1$, $K_2$ and $K_3$ values, as shown in Fig.~\ref{fig:r1_combined} by the dot-dashed curves. Our results show excellent agreement with the explosive transitions from synchronized to incoherent states.

In summary, we have found critical synchronization transitions for generic symmetric natural frequency distributions via a self-consistency approach. Our analysis shows that drifters play a crucial role in controlling synchronization bifurcations, and so the whole natural frequency distribution must be accounted for. This is unlike most other self-consistency analyses, in which the drifters can be neglected, and, hence, bifurcations are determined exclusively by the shape of the distribution at its center. We have shown the efficacy of our methodology for both Lorentzian and Gaussian natural frequency distributions. 

Our methodology can be readily extended to a broad class of possible higher-order interactions, with the only requirement being that the mean-field dynamics can be written in the form (\ref{eq:HO_KM_MF_2}) for some $\gamma$ that encodes the relevant Daido order parameters and their weightings.

\begin{center}
{\bf{Appendix A: Quantifying the effect of drifters}}
\end{center}

Repeating equation (\ref{eq:SC_rk_d}) from the main text, the effect of drifters on the order parameter $r_k$ is
\begin{equation}  \nonumber
r_k^d = \int_{|\omega|>\gamma} \int_{-\pi}^\pi \cos(k\theta) g(\omega) \rho(\theta,\omega) d\theta d\omega.
\end{equation}
We define
\begin{equation}
\tilde{I}_k(\omega) =  \int_{-\pi}^\pi \cos(k\theta) \rho(\theta,\omega) d\theta,
\end{equation}
and will show that this is an odd function if $k$ is odd and an even function if $k$ is even. It then follows that 
\begin{align}  
r_k^d &= \int_{|\omega|>\gamma} \tilde{I}_k(\omega) g(\omega) d\omega \nonumber \\
& =
 \begin{cases} 
0, & k\in \text{odd}, \\ 
2 \int_{\gamma}^\infty \tilde{I}_k(\omega) g(\omega) d\omega, & k\in \text{even},
\end{cases}\label{eq:rk_d}
\end{align}
since $g(\omega)$ is assumed to be even. Returning to $\tilde{I}_k(\omega)$, the integral can be split as
\begin{equation} \label{eq:I_tilde}
\tilde{I}_k(\omega) =  \int_{-\pi}^0 \cos(k\theta) \rho(\theta,\omega) d\theta + \int_{0}^\pi \cos(k\theta) \rho(\theta,\omega) d\theta.
\end{equation}
Focusing on the second integral, after a change of variables $\phi = \theta -\pi$ we obtain
\begin{equation}
\int_{-\pi}^0 \cos(k(\phi + \pi)) \rho(\phi+\pi,\omega) d\phi = \int_{-\pi}^0 (-1)^k \cos(k\phi) \rho(\phi,-\omega) d\phi,
\end{equation}
where we have used the symmetry $\rho(\theta+\pi,-\omega) = \rho(\theta,\omega)$. Substituting back into (\ref{eq:I_tilde}) yields
\begin{equation}
\tilde{I}_k(\omega) = \int_{-\pi}^0 \cos(k\theta)\left( \rho(\theta,\omega) + (-1)^k\rho(\theta,-\omega)\right) d\theta,
\end{equation}
from which it is clear that $\tilde{I}_k(\omega)$ is odd if $k$ is odd and even if $k$ is even. We note that (\ref{eq:SC_r2_d}) from the main text follows directly from (\ref{eq:rk_d}) for $k=2$ after a change of variables $\eta = \omega/\gamma$.

We now derive in full detail the Taylor series expansion (\ref{eq:r2_d_TS}) of $r_2^d$ about $\gamma=0$. We will write the Taylor series as $r_2^d(\gamma) = c_0 + c_1 \gamma + c_2 \gamma^2 + \mathcal{O}(\gamma^3)$. First, since $\int_1^\infty I_\theta(\eta) d\eta = -1/3$ converges, if $g(\omega)$ is bounded (as it generically is), then using the dominated convergence theorem it follows that
\begin{equation} \label{eq:switch_lim_and_int}
\lim_{\gamma \to 0} \int_1^\infty I_\theta(\eta)g(\gamma \eta) d\eta = \int_1^\infty \lim_{\gamma \to 0} I_\theta(\eta)g(\gamma \eta) d\eta = -\frac{g(0)}{3}.
\end{equation}
Hence,
\begin{equation}
c_0 = \lim_{\gamma \to 0} r_2^d = \lim_{\gamma \to 0} 2 \gamma \left(-\frac{g(0)}{3}\right) = 0,
\end{equation}
and so the constant term in the Taylor series is zero. For the linear term in the Taylor series, we can write
\begin{equation}
c_1 = \lim_{\gamma \to 0} \frac{r_2^d(\gamma)}{\gamma} =2 \lim_{\gamma \to 0}  \int_1^\infty I_\theta(\eta) g(\gamma \eta) d\eta =  -\frac{2 g(0)}{3},
\end{equation}
where we have again used (\ref{eq:switch_lim_and_int}). For the quadratic term in the series, we have
\begin{equation} \label{eq:c2_1}
c_2 = \lim_{\gamma \to 0}  \frac{ \gamma\frac{dr_2^d}{d\gamma} - r_2^d(\gamma)}{\gamma^2}.
\end{equation}
From Leibniz rule we obtain 
\begin{align}
\frac{dr_2^d}{d\gamma} &= 2 \int_1^\infty I_\theta(\eta)g(\gamma \eta) d\eta + 2 \gamma \int_1^\infty \eta I_\theta(\eta)g'(\gamma \eta) d\eta \nonumber \\
& = \frac{r_2^d(\gamma)}{\gamma} + 2 \gamma \int_1^\infty \eta I_\theta(\eta)g'(\gamma \eta) d\eta. \label{eq:dr2dgamma}
\end{align}
Substituting (\ref{eq:dr2dgamma}) into (\ref{eq:c2_1}) yields
\begin{equation} \label{eq:c2_2}
c_2 = 2 \lim_{\gamma \to 0}  \int_1^\infty \eta I_\theta(\eta)g'(\gamma \eta) d\eta.
\end{equation}
We note that for Lorentzian and Gaussian frequency distributions, $r_2^d(\gamma)$ can be computed directly, and its Taylor series computed subsequently, as an alternative to using (\ref{eq:c2_2}).

\begin{center}
{\bf{Appendix B: An explicit example showing that the entire frequency distribution must be accounted for}}
\end{center}

Here we use an explicit example to show that unlike the onset of synchronization in the Kuramoto model, which depends only on the properties of $g(\omega)$ at $\omega = 0$, the onset of synchronization in the higher-order model (\ref{eq:HO_KM}) is determined by the entirety of $g(\omega)$. This originates from the inability to interchange the limit and integral in (\ref{eq:c2_2}), and can be physically interpreted as quantifying the effect of drifters, which we have shown to be non-negligible in the case $K_2\neq 0$.

\begin{figure}[tbp]
 \centering
 \includegraphics[width=\columnwidth]{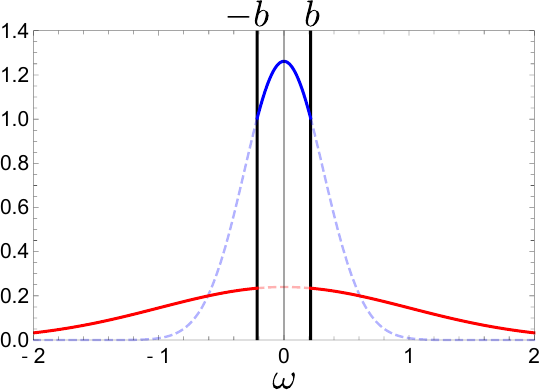}
 \caption{
 The hybrid distribution $g_2(\omega)$ (\ref{eq:hybrid_distro}) for $\sigma^2 = 0.1$, $s=1$ and $b= \sqrt{2} \sigma \erfc^{-1}(1/2) \approx 0.2133$ such that $\int_{-b}^b g_1(\omega) d\omega = 1/2$. The distribution $g_1(\omega)$ is shown in blue and $f(\omega)/\alpha$ is shown in red.
 }
 \label{fig:hybrid_gaussian}
 \end{figure}

We compare two frequency distributions that are identical in the range $\omega \in (-b,b)$, but have different tails. The first distribution is a standard Gaussian distribution with mean zero and variance $\sigma^2$, with PDF $g_1(\omega)$. We show in the main text that the critical values for $g_1(\omega)$ are $K_c = 2\sigma \sqrt{2/\pi}$, $c_2 = 1/(4\sigma^2)$ and $K_{2,3}^* = 2\sigma \sqrt{2/\pi^3}$. As a second distribution, we consider the hybrid PDF
\begin{equation} \label{eq:hybrid_distro}
g_2(\omega) = \begin{cases}
g_1(\omega), & |\omega|<b, \\
f(\omega)/\alpha, & |\omega|>b,
\end{cases}
\end{equation}
where $g_1 \sim \mathcal{N}(0,\sigma^2)$ is the same as our first distribution, $f(\omega) \sim \mathcal{N}(0,s^2)$ is a second Gaussian distribution with a larger variance, and 
\begin{equation}
\alpha = \frac{\erfc\left( D \right)}{\erfc\left( B \right)}
\end{equation}
 is a renormalization constant ensuring that $\int g_2 d\omega = 1$, where $B = b/(\sqrt{2} \sigma)$ and $D = b/(\sqrt{2} s)$. An example of such a distribution $g_2(\omega)$ is shown in Fig.~\ref{fig:hybrid_gaussian} using $\sigma^2 = 0.1$, $s=1$ and $b= \sqrt{2} \sigma \erfc^{-1}(1/2) \approx 0.2133$ such that $\int_{-b}^b g_1(\omega) d\omega = 1/2$.

We compute $r_2^d(\gamma)$ and its Taylor series about $\gamma=0$ explicitly, using (\ref{eq:rk_d}). Since we are interested in the limit $\gamma\to 0$, we compute $r_2^d(\gamma)$ only for $\gamma<b$. For the center part of the distribution, with $|\omega|<b$ and $g_2 = g_1$, we obtain
\begin{widetext}
\begin{equation} \label{eq:hybrid_inner}
2 \int_{\gamma}^b \tilde{I}_2(\omega) g_1(\omega) d\omega = \frac{1}{A^2} \left(\left(A^2-1\right) \left(\erf(B) - \erf(A)\right) + \frac{1}{\sqrt{\pi}} \left( 2B e^{-B^2} + e^{-A^2} \left( \sqrt{\pi} - 2A - 2 \Gamma(3/2,B^2-A^2) \right) \right) \right),
\end{equation}
\end{widetext}
where $A = \gamma/(\sqrt{2}\sigma)$, as in (\ref{eq:r1_Gaussian})-(\ref{eq:r2_Gaussian}) and $\Gamma(a,z)$ denotes the incomplete Gamma function.
The associated Taylor series expansion is
\begin{equation} \label{eq:hybrid_TS_inner}
-\frac{2g_1(0)}{3} \gamma + \frac{1}{4\sigma^2} \left(  \frac{e^{-B^2}}{\sqrt{\pi} B} + \erf(B) \right) \gamma^2 + \mathcal{O}(\gamma^3)
\end{equation}
Here the $\mathcal{O}(\gamma)$ term is identical to that of the standard Gaussian distribution $g_1(\omega)$ (cf. (\ref{eq:r2_d_TS})).
For the tails, with $|\omega|>b$ and $g_2 = f/\alpha$, we obtain
\begin{widetext}
\begin{equation} \label{eq:hybrid_outer}
2 \int_b^\infty \tilde{I}_2(\omega) \frac{f(\omega)}{\alpha } d\omega =\frac{ \erfc(B)}{C^2 \erfc(D)} \left( \left(C^2 - 1\right) \erfc(D) + \frac{2}{\sqrt{\pi}} \left( e^{-C^2} \Gamma(3/2,D^2 - C^2) - De^{-D^2} \right) \right),
\end{equation}
\end{widetext}
where $C = \gamma/(\sqrt{2}s)$.
The corresponding Taylor series expansion for the tails is
\begin{equation} \label{eq:hybrid_TS_outer}
\frac{\erfc(B) \left( 2D\, \Gamma(3/2,D^2) - e^{-D^2} \left(1 + 2D^2 \right)   \right)}{4 \sqrt{\pi} s^2 D \erfc(D)} \gamma^2 + \mathcal{O}(\gamma^4)
\end{equation}
We note that, as expected, the tails do not contribute to the $\mathcal{O}(\gamma)$ term, and only contribute to the higher order terms. 

\begin{figure}[tbp]
\centering
\includegraphics[width=\columnwidth]{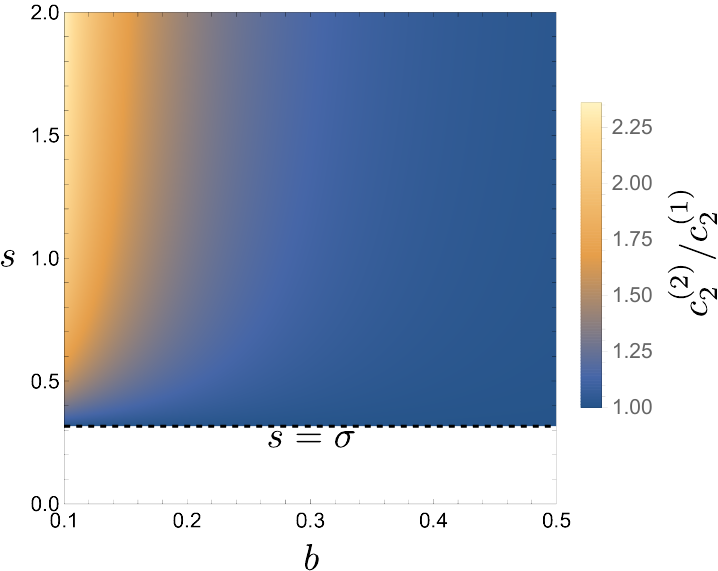}
\caption{
The ratio $c_2^{(2)}/c_2^{(1)}$ of $\mathcal{O}(\gamma^2)$ terms in the Taylor series of $r_2^d(\gamma)$ for $g_1$ and $g_2$ over a range of $b$ and $s$ values, with $\sigma^2 =0.1$ kept constant.
}
\label{fig:c22_over_c21}
\end{figure}

Therefore, $r_2^d(\gamma)$ for the hybrid distribution $g_2(\omega)$ is given by the sum of (\ref{eq:hybrid_inner}) and (\ref{eq:hybrid_outer}). These equations can in principle be solved to find the solution branches for $r_1$ and $r_2$. As discussed in the main text, criticality of the pitchfork bifurcation at $K_1 = K_c$ can be determined from the Taylor series expansion of $r_2^d(\gamma)$ about $\gamma=0$, which, for $g_2$, is given by the sum of (\ref{eq:hybrid_TS_inner}) and (\ref{eq:hybrid_TS_outer}). This Taylor series should be compared to the Taylor series corresponding to the standard Gaussian $g_1(\omega)$, which is given by
\begin{equation}
-\frac{2g_1(0)}{3} \gamma + \frac{1}{4\sigma^2} \gamma^2 + \mathcal{O}(\gamma^3).
\end{equation}
We see that the $\mathcal{O}(\gamma)$ term is the same for both $g_1$ and $g_2$, and depends only on the value of $g_1(\omega)$ at $\omega = 0$, i.e., it is independent of the tails of the distributions. On the other hand, the $\mathcal{O}(\gamma^2)$ terms are markedly different, which will yield different bifurcation values from gradual to explosive synchronization according to (\ref{eq:crit_K2_K3}). Letting $c_2^{(1)} = 1/(4\sigma^2)$ denote the $\mathcal{O}(\gamma^2)$ term corresponding to $g_1(\omega)$, and $c_2^{(2)}$ denote the $\mathcal{O}(\gamma^2)$ term corresponding to $g_2(\omega)$ (summing (\ref{eq:hybrid_TS_inner}) and (\ref{eq:hybrid_TS_outer})), Fig.~\ref{fig:c22_over_c21} shows that for $s>\sigma$, with $\sigma^2 = 0.1$ kept fixed, $c_2^{(2)}/c_2^{(1)}$ takes on a range of values greater than $1$, increasing as the width of the inner region, $b$, approaches zero.

\begin{figure*}[tbp]
\centering
\includegraphics[width=\textwidth]{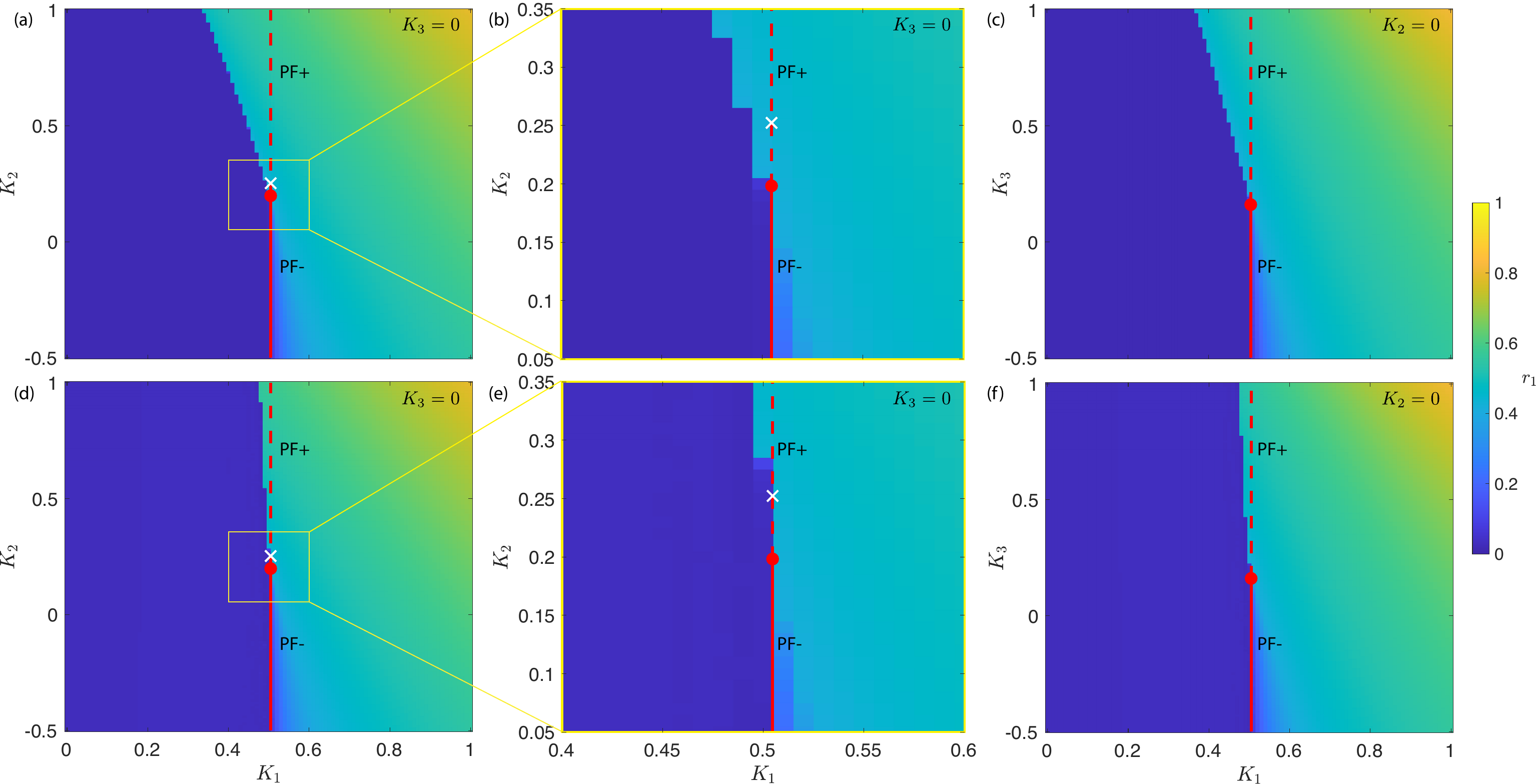}
\caption{
The order parameter $r_1$ for the hybrid distribution $g_2(\omega)$ (\ref{eq:hybrid_distro}) with $\sigma^2 = 0.1$, $s=1$ and $b= \sqrt{2} \sigma \erfc^{-1}(1/2) \approx 0.2133$ (cf. Fig.~\ref{fig:hybrid_gaussian}) for a range of $K_1$, $K_2$ and $K_3$ values using $N=10^3$. As in Fig.~\ref{fig:r1_combined}, the theoretical bifurcation values, based on (\ref{eq:kc}) and (\ref{eq:crit_K2_K3}) are shown in red, with solid representing supercritical pitchfork bifurcation (PF-) and dashed representing subcritical pitchfork bifurcation (PF+). The co-dimension two bifurcation point given by (\ref{eq:kc}) and (\ref{eq:crit_K2_K3}) is shown as a red circle. Top row: Highly synchronized initial condition. Bottom row: Uniformly random initial condition. (a), (d)~$K_3=0$. The white cross shows the theoretical bifurcation point for the standard Gaussian $g_1(\omega)$. (b), (e)~Zoom ins of (a) and (d) near the co-dimension 2 bifurcation. (c), (f)~$K_2=0$.
}
\label{fig:hybrid_r1}
\end{figure*}

To verify our above analytic results, we compare with numerical simulations of the full model (\ref{eq:HO_KM}) using $N=10^3$. We consider fixed $\sigma^2 =0.1$, $s^2=1$, and choose $b$ such that
\begin{equation}
\int_{-b}^b g_1(\omega) d\omega = \frac{1}{2},
\end{equation}
explicitly, $b = \sqrt{2} \sigma \erfc^{-1}(1/2) \approx 0.2133$ (cf. Fig.~\ref{fig:hybrid_gaussian}). Since the shape of both $g_1$ and $g_2$ is identical at $\omega=0$, our analytic results predict that both distributions will have the same critical $K_1 = K_c = 2\sigma \sqrt{2/\pi} \approx 0.5046$, which agrees with the results from numerical simulations shown in Fig.~\ref{fig:hybrid_r1}.  The respective $\mathcal{O}(\gamma^2)$ terms from the Taylor series expansions of $r_2^d(\gamma)$ are $c_2^{(1)} = 5/2$ and $c_2^{(2)} \approx 3.1807$. Therefore, for $K_2 \neq 0$ we expect different critical values at which synchronization transitions from gradual to explosive. For $K_3 = 0$, the critical $K_2$ values are $K_2^{(1)} = 0.2533$ for $g_1$ and $K_2^{(2)} = 0.1991$ for $g_2$. This is confirmed in our numerical results shown in Fig.~\ref{fig:hybrid_r1}(a,b,e,f), where the red circle shows the theoretical bifurcation location for $g_2$ which agrees with the simulations, whereas the bifurcation value for $g_1$, shown as a white cross is not accurate. Moreover, for $K_2=0$, we expect the critical value $K_3 = 0.1606$ at which synchronization transitions from gradual to explosive to be the same for both $g_1$ and $g_2$, since the drifters can be neglected in this case. Our numerical results shown in Fig.~\ref{fig:hybrid_r1}(c,f) verify that the co-dimension 2 bifurcation location is unchanged.


%


\begin{thebibliography}{32}%
\makeatletter
\providecommand \@ifxundefined [1]{%
 \@ifx{#1\undefined}
}%
\providecommand \@ifnum [1]{%
 \ifnum #1\expandafter \@firstoftwo
 \else \expandafter \@secondoftwo
 \fi
}%
\providecommand \@ifx [1]{%
 \ifx #1\expandafter \@firstoftwo
 \else \expandafter \@secondoftwo
 \fi
}%
\providecommand \natexlab [1]{#1}%
\providecommand \enquote  [1]{``#1''}%
\providecommand \bibnamefont  [1]{#1}%
\providecommand \bibfnamefont [1]{#1}%
\providecommand \citenamefont [1]{#1}%
\providecommand \href@noop [0]{\@secondoftwo}%
\providecommand \href [0]{\begingroup \@sanitize@url \@href}%
\providecommand \@href[1]{\@@startlink{#1}\@@href}%
\providecommand \@@href[1]{\endgroup#1\@@endlink}%
\providecommand \@sanitize@url [0]{\catcode `\\12\catcode `\$12\catcode
  `\&12\catcode `\#12\catcode `\^12\catcode `\_12\catcode `\%12\relax}%
\providecommand \@@startlink[1]{}%
\providecommand \@@endlink[0]{}%
\providecommand \url  [0]{\begingroup\@sanitize@url \@url }%
\providecommand \@url [1]{\endgroup\@href {#1}{\urlprefix }}%
\providecommand \urlprefix  [0]{URL }%
\providecommand \Eprint [0]{\href }%
\providecommand \doibase [0]{https://doi.org/}%
\providecommand \selectlanguage [0]{\@gobble}%
\providecommand \bibinfo  [0]{\@secondoftwo}%
\providecommand \bibfield  [0]{\@secondoftwo}%
\providecommand \translation [1]{[#1]}%
\providecommand \BibitemOpen [0]{}%
\providecommand \bibitemStop [0]{}%
\providecommand \bibitemNoStop [0]{.\EOS\space}%
\providecommand \EOS [0]{\spacefactor3000\relax}%
\providecommand \BibitemShut  [1]{\csname bibitem#1\endcsname}%
\let\auto@bib@innerbib\@empty
\bibitem [{\citenamefont {Bhowmik}\ and\ \citenamefont
  {Shanahan}(2012)}]{BhowmikShanahan12}%
  \BibitemOpen
  \bibfield  {author} {\bibinfo {author} {\bibfnamefont {D.}~\bibnamefont
  {Bhowmik}}\ and\ \bibinfo {author} {\bibfnamefont {M.}~\bibnamefont
  {Shanahan}},\ }\bibfield  {title} {\bibinfo {title} {How well do oscillator
  models capture the behaviour of biological neurons?},\ }in\ \href@noop {}
  {\emph {\bibinfo {booktitle} {2012 Int. Joint Conf. Neural Net. (IJCNN)}}}\
  (\bibinfo {year} {2012})\BibitemShut {NoStop}%
\bibitem [{\citenamefont {Bick}\ \emph {et~al.}(2020)\citenamefont {Bick},
  \citenamefont {Goodfellow}, \citenamefont {Laing},\ and\ \citenamefont
  {Martens}}]{BickEtAl2020}%
  \BibitemOpen
  \bibfield  {author} {\bibinfo {author} {\bibfnamefont {C.}~\bibnamefont
  {Bick}}, \bibinfo {author} {\bibfnamefont {M.}~\bibnamefont {Goodfellow}},
  \bibinfo {author} {\bibfnamefont {C.~R.}\ \bibnamefont {Laing}},\ and\
  \bibinfo {author} {\bibfnamefont {E.~A.}\ \bibnamefont {Martens}},\
  }\bibfield  {title} {\bibinfo {title} {Understanding the dynamics of
  biological and neural oscillator networks through exact mean-field
  reductions: a review},\ }\href@noop {} {\bibfield  {journal} {\bibinfo
  {journal} {J. Math. Neurosci.}\ }\textbf {\bibinfo {volume} {10}},\ \bibinfo
  {pages} {1} (\bibinfo {year} {2020})}\BibitemShut {NoStop}%
\bibitem [{\citenamefont {Machowski}\ \emph {et~al.}(2011)\citenamefont
  {Machowski}, \citenamefont {Bialek},\ and\ \citenamefont
  {Bumby}}]{MachowskiEtAl2011}%
  \BibitemOpen
  \bibfield  {author} {\bibinfo {author} {\bibfnamefont {J.}~\bibnamefont
  {Machowski}}, \bibinfo {author} {\bibfnamefont {J.~W.}\ \bibnamefont
  {Bialek}},\ and\ \bibinfo {author} {\bibfnamefont {J.}~\bibnamefont
  {Bumby}},\ }\href@noop {} {\emph {\bibinfo {title} {Power system dynamics:
  stability and control}}}\ (\bibinfo  {publisher} {John Wiley \& Sons},\
  \bibinfo {year} {2011})\BibitemShut {NoStop}%
\bibitem [{\citenamefont {Nishikawa}\ and\ \citenamefont
  {Motter}(2015)}]{NishikawaMotter2015}%
  \BibitemOpen
  \bibfield  {author} {\bibinfo {author} {\bibfnamefont {T.}~\bibnamefont
  {Nishikawa}}\ and\ \bibinfo {author} {\bibfnamefont {A.~E.}\ \bibnamefont
  {Motter}},\ }\bibfield  {title} {\bibinfo {title} {Comparative analysis of
  existing models for power-grid synchronization},\ }\bibfield  {journal}
  {\bibinfo  {journal} {New J. Phys.}\ }\textbf {\bibinfo {volume} {17}},\
  \href {https://doi.org/10.1088/1367-2630/17/1/015012}
  {10.1088/1367-2630/17/1/015012} (\bibinfo {year} {2015})\BibitemShut
  {NoStop}%
\bibitem [{\citenamefont {Filatrella}\ \emph {et~al.}(2008)\citenamefont
  {Filatrella}, \citenamefont {Nielsen},\ and\ \citenamefont
  {Pedersen}}]{FilatrellaEtAl08}%
  \BibitemOpen
  \bibfield  {author} {\bibinfo {author} {\bibfnamefont {G.}~\bibnamefont
  {Filatrella}}, \bibinfo {author} {\bibfnamefont {A.~H.}\ \bibnamefont
  {Nielsen}},\ and\ \bibinfo {author} {\bibfnamefont {N.~F.}\ \bibnamefont
  {Pedersen}},\ }\bibfield  {title} {\bibinfo {title} {Analysis of a power grid
  using a {K}uramoto-like model},\ }\href@noop {} {\bibfield  {journal}
  {\bibinfo  {journal} {Eur. Phys. J. B}\ }\textbf {\bibinfo {volume} {61}}
  (\bibinfo {year} {2008})}\BibitemShut {NoStop}%
\bibitem [{\citenamefont {Sch\"{a}fer}\ and\ \citenamefont
  {Yalcin}(2019)}]{SchaferYalcin2019}%
  \BibitemOpen
  \bibfield  {author} {\bibinfo {author} {\bibfnamefont {B.}~\bibnamefont
  {Sch\"{a}fer}}\ and\ \bibinfo {author} {\bibfnamefont {G.~C.}\ \bibnamefont
  {Yalcin}},\ }\bibfield  {title} {\bibinfo {title} {Dynamical modeling of
  cascading failures in the {T}urkish power grid},\ }\bibfield  {journal}
  {\bibinfo  {journal} {Chaos}\ }\textbf {\bibinfo {volume} {29}},\ \href
  {https://doi.org/10.1063/1.5110974} {10.1063/1.5110974} (\bibinfo {year}
  {2019})\BibitemShut {NoStop}%
\bibitem [{\citenamefont {Petri}\ \emph {et~al.}(2014)\citenamefont {Petri},
  \citenamefont {Expert}, \citenamefont {Turkheimer}, \citenamefont
  {Carhart-Harris}, \citenamefont {Nutt}, \citenamefont {Hellyer},\ and\
  \citenamefont {Vaccarino}}]{PetriEtAl2014}%
  \BibitemOpen
  \bibfield  {author} {\bibinfo {author} {\bibfnamefont {G.}~\bibnamefont
  {Petri}}, \bibinfo {author} {\bibfnamefont {P.}~\bibnamefont {Expert}},
  \bibinfo {author} {\bibfnamefont {F.}~\bibnamefont {Turkheimer}}, \bibinfo
  {author} {\bibfnamefont {R.}~\bibnamefont {Carhart-Harris}}, \bibinfo
  {author} {\bibfnamefont {D.}~\bibnamefont {Nutt}}, \bibinfo {author}
  {\bibfnamefont {P.~J.}\ \bibnamefont {Hellyer}},\ and\ \bibinfo {author}
  {\bibfnamefont {F.}~\bibnamefont {Vaccarino}},\ }\bibfield  {title} {\bibinfo
  {title} {Homological scaffolds of brain functional networks},\ }\href
  {https://doi.org/10.1098/rsif.2014.0873} {\bibfield  {journal} {\bibinfo
  {journal} {J. Roy. Soc. Interface}\ }\textbf {\bibinfo {volume} {11}},\
  \bibinfo {pages} {20140873} (\bibinfo {year} {2014})}\BibitemShut {NoStop}%
\bibitem [{\citenamefont {Lord}\ \emph {et~al.}(2016)\citenamefont {Lord},
  \citenamefont {Expert}, \citenamefont {Fernandes}, \citenamefont {Petri},
  \citenamefont {Van~Hartevelt}, \citenamefont {Vaccarino}, \citenamefont
  {Deco}, \citenamefont {Turkheimer},\ and\ \citenamefont
  {Kringelbach}}]{LordEtAl2016}%
  \BibitemOpen
  \bibfield  {author} {\bibinfo {author} {\bibfnamefont {L.-D.}\ \bibnamefont
  {Lord}}, \bibinfo {author} {\bibfnamefont {P.}~\bibnamefont {Expert}},
  \bibinfo {author} {\bibfnamefont {H.~M.}\ \bibnamefont {Fernandes}}, \bibinfo
  {author} {\bibfnamefont {G.}~\bibnamefont {Petri}}, \bibinfo {author}
  {\bibfnamefont {T.~J.}\ \bibnamefont {Van~Hartevelt}}, \bibinfo {author}
  {\bibfnamefont {F.}~\bibnamefont {Vaccarino}}, \bibinfo {author}
  {\bibfnamefont {G.}~\bibnamefont {Deco}}, \bibinfo {author} {\bibfnamefont
  {F.}~\bibnamefont {Turkheimer}},\ and\ \bibinfo {author} {\bibfnamefont
  {M.~L.}\ \bibnamefont {Kringelbach}},\ }\bibfield  {title} {\bibinfo {title}
  {Insights into brain architectures from the homological scaffolds of
  functional connectivity networks},\ }\bibfield  {journal} {\bibinfo
  {journal} {Front. Syst. Neurosci.}\ }\textbf {\bibinfo {volume} {10}},\ \href
  {https://doi.org/10.3389/fnsys.2016.00085} {10.3389/fnsys.2016.00085}
  (\bibinfo {year} {2016})\BibitemShut {NoStop}%
\bibitem [{\citenamefont {Battiston}\ \emph {et~al.}(2021)\citenamefont
  {Battiston}, \citenamefont {Amico}, \citenamefont {Barrat}, \citenamefont
  {Bianconi}, \citenamefont {Ferraz~de Arruda}, \citenamefont {Franceschiello},
  \citenamefont {Iacopini}, \citenamefont {K{\'e}fi}, \citenamefont {Latora},
  \citenamefont {Moreno}, \citenamefont {Murray}, \citenamefont {Peixoto},
  \citenamefont {Vaccarino},\ and\ \citenamefont {Petri}}]{BattistonEtAl2021}%
  \BibitemOpen
  \bibfield  {author} {\bibinfo {author} {\bibfnamefont {F.}~\bibnamefont
  {Battiston}}, \bibinfo {author} {\bibfnamefont {E.}~\bibnamefont {Amico}},
  \bibinfo {author} {\bibfnamefont {A.}~\bibnamefont {Barrat}}, \bibinfo
  {author} {\bibfnamefont {G.}~\bibnamefont {Bianconi}}, \bibinfo {author}
  {\bibfnamefont {G.}~\bibnamefont {Ferraz~de Arruda}}, \bibinfo {author}
  {\bibfnamefont {B.}~\bibnamefont {Franceschiello}}, \bibinfo {author}
  {\bibfnamefont {I.}~\bibnamefont {Iacopini}}, \bibinfo {author}
  {\bibfnamefont {S.}~\bibnamefont {K{\'e}fi}}, \bibinfo {author}
  {\bibfnamefont {V.}~\bibnamefont {Latora}}, \bibinfo {author} {\bibfnamefont
  {Y.}~\bibnamefont {Moreno}}, \bibinfo {author} {\bibfnamefont {M.~M.}\
  \bibnamefont {Murray}}, \bibinfo {author} {\bibfnamefont {T.~P.}\
  \bibnamefont {Peixoto}}, \bibinfo {author} {\bibfnamefont {F.}~\bibnamefont
  {Vaccarino}},\ and\ \bibinfo {author} {\bibfnamefont {G.}~\bibnamefont
  {Petri}},\ }\bibfield  {title} {\bibinfo {title} {The physics of higher-order
  interactions in complex systems},\ }\href
  {https://doi.org/10.1038/s41567-021-01371-4} {\bibfield  {journal} {\bibinfo
  {journal} {Nat. Phys.}\ }\textbf {\bibinfo {volume} {17}},\ \bibinfo {pages}
  {1093} (\bibinfo {year} {2021})}\BibitemShut {NoStop}%
\bibitem [{\citenamefont {Majhi}\ \emph {et~al.}(2022)\citenamefont {Majhi},
  \citenamefont {Perc},\ and\ \citenamefont {Ghosh}}]{MajhiEtAl2022}%
  \BibitemOpen
  \bibfield  {author} {\bibinfo {author} {\bibfnamefont {S.}~\bibnamefont
  {Majhi}}, \bibinfo {author} {\bibfnamefont {M.}~\bibnamefont {Perc}},\ and\
  \bibinfo {author} {\bibfnamefont {D.}~\bibnamefont {Ghosh}},\ }\bibfield
  {title} {\bibinfo {title} {Dynamics on higher-order networks: a review},\
  }\href {https://doi.org/10.1098/rsif.2022.0043} {\bibfield  {journal}
  {\bibinfo  {journal} {Journal of The Royal Society Interface}\ }\textbf
  {\bibinfo {volume} {19}},\ \bibinfo {pages} {20220043} (\bibinfo {year}
  {2022})}\BibitemShut {NoStop}%
\bibitem [{\citenamefont {Battiston}\ \emph {et~al.}(2020)\citenamefont
  {Battiston}, \citenamefont {Cencetti}, \citenamefont {Iacopini},
  \citenamefont {Latora}, \citenamefont {Lucas}, \citenamefont {Patania},
  \citenamefont {Young},\ and\ \citenamefont {Petri}}]{BattistonEtAl2020}%
  \BibitemOpen
  \bibfield  {author} {\bibinfo {author} {\bibfnamefont {F.}~\bibnamefont
  {Battiston}}, \bibinfo {author} {\bibfnamefont {G.}~\bibnamefont {Cencetti}},
  \bibinfo {author} {\bibfnamefont {I.}~\bibnamefont {Iacopini}}, \bibinfo
  {author} {\bibfnamefont {V.}~\bibnamefont {Latora}}, \bibinfo {author}
  {\bibfnamefont {M.}~\bibnamefont {Lucas}}, \bibinfo {author} {\bibfnamefont
  {A.}~\bibnamefont {Patania}}, \bibinfo {author} {\bibfnamefont {J.-G.}\
  \bibnamefont {Young}},\ and\ \bibinfo {author} {\bibfnamefont
  {G.}~\bibnamefont {Petri}},\ }\bibfield  {title} {\bibinfo {title} {Networks
  beyond pairwise interactions: Structure and dynamics},\ }\href
  {https://doi.org/https://doi.org/10.1016/j.physrep.2020.05.004} {\bibfield
  {journal} {\bibinfo  {journal} {Physics Reports}\ }\textbf {\bibinfo {volume}
  {874}},\ \bibinfo {pages} {1} (\bibinfo {year} {2020})},\ \bibinfo {note}
  {networks beyond pairwise interactions: Structure and dynamics}\BibitemShut
  {NoStop}%
\bibitem [{\citenamefont {Mill\'an}\ \emph {et~al.}(2020)\citenamefont
  {Mill\'an}, \citenamefont {Torres},\ and\ \citenamefont
  {Bianconi}}]{MillanEtAl2020}%
  \BibitemOpen
  \bibfield  {author} {\bibinfo {author} {\bibfnamefont {A.~P.}\ \bibnamefont
  {Mill\'an}}, \bibinfo {author} {\bibfnamefont {J.~J.}\ \bibnamefont
  {Torres}},\ and\ \bibinfo {author} {\bibfnamefont {G.}~\bibnamefont
  {Bianconi}},\ }\bibfield  {title} {\bibinfo {title} {Explosive higher-order
  kuramoto dynamics on simplicial complexes},\ }\href
  {https://doi.org/10.1103/PhysRevLett.124.218301} {\bibfield  {journal}
  {\bibinfo  {journal} {Phys. Rev. Lett.}\ }\textbf {\bibinfo {volume} {124}},\
  \bibinfo {pages} {218301} (\bibinfo {year} {2020})}\BibitemShut {NoStop}%
\bibitem [{\citenamefont {Skardal}\ and\ \citenamefont
  {Arenas}(2020)}]{SkardalArenas2020}%
  \BibitemOpen
  \bibfield  {author} {\bibinfo {author} {\bibfnamefont {P.~S.}\ \bibnamefont
  {Skardal}}\ and\ \bibinfo {author} {\bibfnamefont {A.}~\bibnamefont
  {Arenas}},\ }\bibfield  {title} {\bibinfo {title} {Higher order interactions
  in complex networks of phase oscillators promote abrupt synchronization
  switching},\ }\href {https://doi.org/10.1038/s42005-020-00485-0} {\bibfield
  {journal} {\bibinfo  {journal} {Commun. Phys.}\ }\textbf {\bibinfo {volume}
  {3}},\ \bibinfo {pages} {218} (\bibinfo {year} {2020})}\BibitemShut {NoStop}%
\bibitem [{\citenamefont {Adhikari}\ \emph {et~al.}(2023)\citenamefont
  {Adhikari}, \citenamefont {Restrepo},\ and\ \citenamefont
  {Skardal}}]{AdhikariEtAl2023}%
  \BibitemOpen
  \bibfield  {author} {\bibinfo {author} {\bibfnamefont {S.}~\bibnamefont
  {Adhikari}}, \bibinfo {author} {\bibfnamefont {J.~G.}\ \bibnamefont
  {Restrepo}},\ and\ \bibinfo {author} {\bibfnamefont {P.~S.}\ \bibnamefont
  {Skardal}},\ }\bibfield  {title} {\bibinfo {title} {{Synchronization of phase
  oscillators on complex hypergraphs}},\ }\bibfield  {journal} {\bibinfo
  {journal} {Chaos}\ }\textbf {\bibinfo {volume} {33}},\ \href
  {https://doi.org/10.1063/5.0116747} {10.1063/5.0116747} (\bibinfo {year}
  {2023}),\ \bibinfo {note} {033116}\BibitemShut {NoStop}%
\bibitem [{\citenamefont {Zhang}\ \emph {et~al.}(2023)\citenamefont {Zhang},
  \citenamefont {Lucas},\ and\ \citenamefont {Battiston}}]{ZhangEtAl2023}%
  \BibitemOpen
  \bibfield  {author} {\bibinfo {author} {\bibfnamefont {Y.}~\bibnamefont
  {Zhang}}, \bibinfo {author} {\bibfnamefont {M.}~\bibnamefont {Lucas}},\ and\
  \bibinfo {author} {\bibfnamefont {F.}~\bibnamefont {Battiston}},\ }\bibfield
  {title} {\bibinfo {title} {Higher-order interactions shape collective
  dynamics differently in hypergraphs and simplicial complexes},\ }\href
  {https://doi.org/10.1038/s41467-023-37190-9} {\bibfield  {journal} {\bibinfo
  {journal} {Nat. Commun.}\ }\textbf {\bibinfo {volume} {14}},\ \bibinfo
  {pages} {1605} (\bibinfo {year} {2023})}\BibitemShut {NoStop}%
\bibitem [{\citenamefont {Gao}\ \emph {et~al.}(2023)\citenamefont {Gao},
  \citenamefont {Ghosh}, \citenamefont {Harrington}, \citenamefont {Restrepo},\
  and\ \citenamefont {Taylor}}]{GaoEtAl2023}%
  \BibitemOpen
  \bibfield  {author} {\bibinfo {author} {\bibfnamefont {Z.}~\bibnamefont
  {Gao}}, \bibinfo {author} {\bibfnamefont {D.}~\bibnamefont {Ghosh}}, \bibinfo
  {author} {\bibfnamefont {H.~A.}\ \bibnamefont {Harrington}}, \bibinfo
  {author} {\bibfnamefont {J.~G.}\ \bibnamefont {Restrepo}},\ and\ \bibinfo
  {author} {\bibfnamefont {D.}~\bibnamefont {Taylor}},\ }\bibfield  {title}
  {\bibinfo {title} {{Dynamics on networks with higher-order interactions}},\
  }\bibfield  {journal} {\bibinfo  {journal} {Chaos}\ }\textbf {\bibinfo
  {volume} {33}},\ \href {https://doi.org/10.1063/5.0151265}
  {10.1063/5.0151265} (\bibinfo {year} {2023}),\ \bibinfo {note}
  {040401}\BibitemShut {NoStop}%
\bibitem [{\citenamefont {Ott}\ and\ \citenamefont
  {Antonsen}(2008)}]{OttAntonsen08}%
  \BibitemOpen
  \bibfield  {author} {\bibinfo {author} {\bibfnamefont {E.}~\bibnamefont
  {Ott}}\ and\ \bibinfo {author} {\bibfnamefont {T.~M.}\ \bibnamefont
  {Antonsen}},\ }\bibfield  {title} {\bibinfo {title} {Low dimensional behavior
  of large systems of globally coupled oscillators},\ }\href
  {https://doi.org/10.1063/1.2930766} {\bibfield  {journal} {\bibinfo
  {journal} {Chaos}\ }\textbf {\bibinfo {volume} {18}},\ \bibinfo {pages}
  {037113, 6} (\bibinfo {year} {2008})}\BibitemShut {NoStop}%
\bibitem [{\citenamefont {Ott}\ and\ \citenamefont
  {Antonsen}(2009)}]{OttAntonsen09}%
  \BibitemOpen
  \bibfield  {author} {\bibinfo {author} {\bibfnamefont {E.}~\bibnamefont
  {Ott}}\ and\ \bibinfo {author} {\bibfnamefont {T.~M.}\ \bibnamefont
  {Antonsen}},\ }\bibfield  {title} {\bibinfo {title} {Long time evolution of
  phase oscillator systems},\ }\href {https://doi.org/10.1063/1.3136851}
  {\bibfield  {journal} {\bibinfo  {journal} {Chaos}\ }\textbf {\bibinfo
  {volume} {19}},\ \bibinfo {pages} {023117} (\bibinfo {year}
  {2009})}\BibitemShut {NoStop}%
\bibitem [{\citenamefont {Kuramoto}(1984)}]{Kuramoto84}%
  \BibitemOpen
  \bibfield  {author} {\bibinfo {author} {\bibfnamefont {Y.}~\bibnamefont
  {Kuramoto}},\ }\href {https://doi.org/10.1007/978-3-642-69689-3} {\emph
  {\bibinfo {title} {Chemical {O}scillations, {W}aves, and {T}urbulence}}}\
  (\bibinfo  {publisher} {Springer-Verlag},\ \bibinfo {address} {Berlin},\
  \bibinfo {year} {1984})\BibitemShut {NoStop}%
\bibitem [{\citenamefont {Strogatz}(2000)}]{Strogatz00}%
  \BibitemOpen
  \bibfield  {author} {\bibinfo {author} {\bibfnamefont {S.~H.}\ \bibnamefont
  {Strogatz}},\ }\bibfield  {title} {\bibinfo {title} {From {K}uramoto to
  {C}rawford: {E}xploring the onset of synchronization in populations of
  coupled oscillators},\ }\bibfield  {journal} {\bibinfo  {journal} {Physica
  D}\ }\textbf {\bibinfo {volume} {143}},\ \href
  {https://doi.org/10.1016/S0167-2789(00)00094-4}
  {10.1016/S0167-2789(00)00094-4} (\bibinfo {year} {2000})\BibitemShut
  {NoStop}%
\bibitem [{\citenamefont {Le\'on}\ and\ \citenamefont
  {Paz\'o}(2019)}]{LeonPazo2019}%
  \BibitemOpen
  \bibfield  {author} {\bibinfo {author} {\bibfnamefont {I.}~\bibnamefont
  {Le\'on}}\ and\ \bibinfo {author} {\bibfnamefont {D.}~\bibnamefont
  {Paz\'o}},\ }\bibfield  {title} {\bibinfo {title} {Phase reduction beyond the
  first order: The case of the mean-field complex {G}inzburg-{L}andau
  equation},\ }\href {https://doi.org/10.1103/PhysRevE.100.012211} {\bibfield
  {journal} {\bibinfo  {journal} {Phys. Rev. E}\ }\textbf {\bibinfo {volume}
  {100}},\ \bibinfo {pages} {012211} (\bibinfo {year} {2019})}\BibitemShut
  {NoStop}%
\bibitem [{\citenamefont {Ashwin}\ and\ \citenamefont
  {Rodrigues}(2016)}]{AshwinRodrigues2016}%
  \BibitemOpen
  \bibfield  {author} {\bibinfo {author} {\bibfnamefont {P.}~\bibnamefont
  {Ashwin}}\ and\ \bibinfo {author} {\bibfnamefont {A.}~\bibnamefont
  {Rodrigues}},\ }\bibfield  {title} {\bibinfo {title} {Hopf normal form with
  sn symmetry and reduction to systems of nonlinearly coupled phase
  oscillators},\ }\href
  {https://doi.org/https://doi.org/10.1016/j.physd.2016.02.009} {\bibfield
  {journal} {\bibinfo  {journal} {Physica D}\ }\textbf {\bibinfo {volume}
  {325}},\ \bibinfo {pages} {14} (\bibinfo {year} {2016})}\BibitemShut
  {NoStop}%
\bibitem [{\citenamefont {Wang}\ \emph
  {et~al.}(2021{\natexlab{a}})\citenamefont {Wang}, \citenamefont {Zheng},\
  and\ \citenamefont {Xu}}]{WangEtAl2021}%
  \BibitemOpen
  \bibfield  {author} {\bibinfo {author} {\bibfnamefont {X.}~\bibnamefont
  {Wang}}, \bibinfo {author} {\bibfnamefont {Z.}~\bibnamefont {Zheng}},\ and\
  \bibinfo {author} {\bibfnamefont {C.}~\bibnamefont {Xu}},\ }\bibfield
  {title} {\bibinfo {title} {Collective dynamics of phase oscillator
  populations with three-body interactions},\ }\href
  {https://doi.org/10.1103/PhysRevE.104.054208} {\bibfield  {journal} {\bibinfo
   {journal} {Phys. Rev. E}\ }\textbf {\bibinfo {volume} {104}},\ \bibinfo
  {pages} {054208} (\bibinfo {year} {2021}{\natexlab{a}})}\BibitemShut
  {NoStop}%
\bibitem [{\citenamefont {Xu}\ \emph {et~al.}(2023)\citenamefont {Xu},
  \citenamefont {Zhai}, \citenamefont {Wu}, \citenamefont {Zheng},\ and\
  \citenamefont {Guan}}]{XuEtAl2023}%
  \BibitemOpen
  \bibfield  {author} {\bibinfo {author} {\bibfnamefont {C.}~\bibnamefont
  {Xu}}, \bibinfo {author} {\bibfnamefont {Y.}~\bibnamefont {Zhai}}, \bibinfo
  {author} {\bibfnamefont {Y.}~\bibnamefont {Wu}}, \bibinfo {author}
  {\bibfnamefont {Z.}~\bibnamefont {Zheng}},\ and\ \bibinfo {author}
  {\bibfnamefont {S.}~\bibnamefont {Guan}},\ }\bibfield  {title} {\bibinfo
  {title} {Enhanced explosive synchronization in heterogeneous oscillator
  populations with higher-order interactions},\ }\href
  {https://doi.org/https://doi.org/10.1016/j.chaos.2023.113343} {\bibfield
  {journal} {\bibinfo  {journal} {Chaos, Solitons \& Fractals}\ }\textbf
  {\bibinfo {volume} {170}},\ \bibinfo {pages} {113343} (\bibinfo {year}
  {2023})}\BibitemShut {NoStop}%
\bibitem [{\citenamefont {Sabhahit}\ \emph {et~al.}(2023)\citenamefont
  {Sabhahit}, \citenamefont {Khurd},\ and\ \citenamefont
  {Jalan}}]{Sabhahit2023}%
  \BibitemOpen
  \bibfield  {author} {\bibinfo {author} {\bibfnamefont {N.~G.}\ \bibnamefont
  {Sabhahit}}, \bibinfo {author} {\bibfnamefont {A.~S.}\ \bibnamefont
  {Khurd}},\ and\ \bibinfo {author} {\bibfnamefont {S.}~\bibnamefont {Jalan}},\
  }\bibfield  {title} {\bibinfo {title} {Self-consistent method for {K}uramoto
  oscillators with inertia having higher-order interactions},\ }\href@noop {}
  {\bibfield  {journal} {\bibinfo  {journal} {arXiv preprint arXiv:2303.08363}\
  } (\bibinfo {year} {2023})}\BibitemShut {NoStop}%
\bibitem [{\citenamefont {Wang}\ \emph
  {et~al.}(2021{\natexlab{b}})\citenamefont {Wang}, \citenamefont {Xu},\ and\
  \citenamefont {Zheng}}]{WangEtAl2021b}%
  \BibitemOpen
  \bibfield  {author} {\bibinfo {author} {\bibfnamefont {X.}~\bibnamefont
  {Wang}}, \bibinfo {author} {\bibfnamefont {C.}~\bibnamefont {Xu}},\ and\
  \bibinfo {author} {\bibfnamefont {Z.}~\bibnamefont {Zheng}},\ }\bibfield
  {title} {\bibinfo {title} {Phase transition and scaling in kuramoto model
  with high-order coupling},\ }\href
  {https://doi.org/10.1007/s11071-021-06268-8} {\bibfield  {journal} {\bibinfo
  {journal} {Nonlinear Dynamics}\ }\textbf {\bibinfo {volume} {103}},\ \bibinfo
  {pages} {2721} (\bibinfo {year} {2021}{\natexlab{b}})}\BibitemShut {NoStop}%
\bibitem [{\citenamefont {Xu}\ \emph {et~al.}(2020)\citenamefont {Xu},
  \citenamefont {Wang},\ and\ \citenamefont {Skardal}}]{XuEtAl2020}%
  \BibitemOpen
  \bibfield  {author} {\bibinfo {author} {\bibfnamefont {C.}~\bibnamefont
  {Xu}}, \bibinfo {author} {\bibfnamefont {X.}~\bibnamefont {Wang}},\ and\
  \bibinfo {author} {\bibfnamefont {P.~S.}\ \bibnamefont {Skardal}},\
  }\bibfield  {title} {\bibinfo {title} {Bifurcation analysis and structural
  stability of simplicial oscillator populations},\ }\href
  {https://doi.org/10.1103/PhysRevResearch.2.023281} {\bibfield  {journal}
  {\bibinfo  {journal} {Phys. Rev. Res.}\ }\textbf {\bibinfo {volume} {2}},\
  \bibinfo {pages} {023281} (\bibinfo {year} {2020})}\BibitemShut {NoStop}%
\bibitem [{\citenamefont {Xu}\ and\ \citenamefont
  {Skardal}(2021)}]{XuEtAl2021}%
  \BibitemOpen
  \bibfield  {author} {\bibinfo {author} {\bibfnamefont {C.}~\bibnamefont
  {Xu}}\ and\ \bibinfo {author} {\bibfnamefont {P.~S.}\ \bibnamefont
  {Skardal}},\ }\bibfield  {title} {\bibinfo {title} {Spectrum of extensive
  multiclusters in the kuramoto model with higher-order interactions},\ }\href
  {https://doi.org/10.1103/PhysRevResearch.3.013013} {\bibfield  {journal}
  {\bibinfo  {journal} {Phys. Rev. Res.}\ }\textbf {\bibinfo {volume} {3}},\
  \bibinfo {pages} {013013} (\bibinfo {year} {2021})}\BibitemShut {NoStop}%
\bibitem [{\citenamefont {Daido}(1996)}]{Daido1996}%
  \BibitemOpen
  \bibfield  {author} {\bibinfo {author} {\bibfnamefont {H.}~\bibnamefont
  {Daido}},\ }\bibfield  {title} {\bibinfo {title} {Multibranch entrainment and
  scaling in large populations of coupled oscillators},\ }\href
  {https://doi.org/10.1103/PhysRevLett.77.1406} {\bibfield  {journal} {\bibinfo
   {journal} {Phys. Rev. Lett.}\ }\textbf {\bibinfo {volume} {77}},\ \bibinfo
  {pages} {1406} (\bibinfo {year} {1996})}\BibitemShut {NoStop}%
\bibitem [{Note1()}]{Note1}%
  \BibitemOpen
  \bibinfo {note} {See Appendix A for full
  detail.}\BibitemShut {Stop}%
\bibitem [{\citenamefont {Kuehn}\ and\ \citenamefont
  {Bick}(2021)}]{KuehnBick2021}%
  \BibitemOpen
  \bibfield  {author} {\bibinfo {author} {\bibfnamefont {C.}~\bibnamefont
  {Kuehn}}\ and\ \bibinfo {author} {\bibfnamefont {C.}~\bibnamefont {Bick}},\
  }\bibfield  {title} {\bibinfo {title} {A universal route to explosive
  phenomena},\ }\href {https://doi.org/10.1126/sciadv.abe3824} {\bibfield
  {journal} {\bibinfo  {journal} {Science Advances}\ }\textbf {\bibinfo
  {volume} {7}},\ \bibinfo {pages} {eabe3824} (\bibinfo {year}
  {2021})}\BibitemShut {NoStop}%
\bibitem [{Note2()}]{Note2}%
  \BibitemOpen
  \bibinfo {note} {See Appendix A for full
  details.}\BibitemShut {Stop}%
\end{thebibliography}

\end{document}